# Influence of particle size and agglomeration in Solid Oxide Fuel Cell cathodes using manganite nanoparticles

Hernán Martinelli[1], Diego G. Lamas[1,2,3], Ana G. Leyva[1,2], Joaquín Sacanell[1,4*]

[1] Departamento de Física de la Materia Condensada, Gerencia de Investigación y Aplicaciones, Centro Atómico Constituyentes, Comisión Nacional de Energía Atómica. San Martín, Pcia. de Buenos Aires, Argentina.
[2] Escuela de Ciencia y Tecnología, Universidad Nacional de Gral. San Martín, San Martín, Pcia. de Buenos Aires, Argentina.
[3] CONICET, Buenos Aires, Argentina.
[4] Instituto de Nanociencia y Nanotecnología (INN), CNEA-CONICET, Buenos Aires, Argentina.

* corresponding author: sacanell@tandar.cnea.gov.ar

In this work we studied the influence of particle size and agglomeration in the performance of solid oxide fuel cell cathodes made with nanoparticles of $La_{0.8}Sr_{0.2}MnO_3$.
We followed two synthesis routes based on the Liquid Mix method. In both procedures we introduced additional reagents in order to separated the manganite particles. We evaluated cathodic performance by Electrochemical Impedance Spectroscopy in symmetrical (CATHODE/ELECTROLYTE/CATHODE) cells. Particle size was tuned by the temperature used for cathode sintering.
Our results show that deagglomeration of the particles, serves to improve the cathodes performance. However, the dependence of the performance with the size of the particles is not clear, as different trends were obtained for each synthesis route.
As a common feature, the cathodes with the lowest area specific resistance are the ones sintered at the largest temperature. This result indicates that an additional factor related with the quality of the cathode/electrolyte sintering, is superimposed with the influence of particle size, however further work is needed to clarify this issue.
The enhancement obtained by deagglomeration suggest that the use of this kind of methods deserved to be considered to develop high performance electrodes for solid oxide fuel cells.

## 1 Introduction

Solid Oxide Fuel Cells (SOFCs) have risen as one of the most efficient energy conversion technology devices, due to their extremely high efficiencies and the possibility of using different fuels such as hydrocarbons or hydrogen [1-3]. The SOFC cathode material must be an electronic conductor (EC) to admit oxygen reduction reaction (ORR) on its surface. An example of these materials are the $La_{1-x}Sr_xMnO_3$ manganites with perovskite structure, particularly with x=0.2 (LSM).[4,5] These materials must have a porous morphology in order to increase the ORR sites, called triple phase boundary (TPB) points and defined by the region where gas, electrolyte and cathode meet.[4-7] Another improvement can be obtained if cathodes are made of mixed electronic and ionic conductors (MIECs), such as $(LaSr)(CoFe)O_3$.[8,9] In that case, porosity is not necessary because oxygen ions can travel through the cathode, however it is beneficial as on a porous morphology, the amount of reaction points increases.[10]

The constant search for new and better materials for SOFCs, is driven by the need to increase performance and to reduce the typically high temperatures of operation of these devices (~1000°C), which is responsible for the degradation of the cell components. Hence, significant efforts have been made in order to obtain an intermediate temperature (IT) SOFC, with novel materials and structures capable of operate below 700ºC.[10-14] Recently, with the development of nanoscience, well-known materials have been revisited due to new properties that arise at this scale. Several examples of improvements could be found in MIECs, particularly in nanotubes of LaSr cobaltites and cobalto-ferrites [14,15]. In contrast, nanostructured EC cathodes have not been widely studied yet, even though their interest has been recognized [16]. A promising material for these studies is nanostructured LSM, which is one of the best known compounds for SOFC cathodes. Studies of microcrystalline LSM by the isotope exchange depth profiling technique using SIMS, have shown that diffusion of oxygen is much faster along grain boundaries than through the bulk.[17,18] Also, recent studies on LSM thin films and nanopowders have reached to the same conclusion and, in addition, a faster oxygen exchange kinetics has been observed.[19,20] Although the appearance of oxide ion conduction was very recently reported in LSM SOFC cathodes, early studies of LSM [21] ascribed the large electrocatalytic activity of LSM to the generation of oxygen vacancies that are mobile enough to carry oxide ions from the surface of the electrode to the electrolyte. Also, cathodes with LSM nanotubes were developed [22], which showed that oxide ion conduction through grain boundaries, can account for the observed electrochemical impedance spectroscopy data. However, due to the synthesis procedure, the nanostructure consisted in nanoparticles agglomerated on the surface of tubular structures of sub-µm diameter and thus, not the whole surface of the nanoparticles is exposed to the atmosphere.
In this work, two chemical techniques based in Liquid Mix have been developed in order to avoid the agglomeration particles in the cathodes. With the resulting materials as precursors, the goal was to obtain nanostructured cathodes and to study their electrochemical properties as a function of particle size and agglomeration.

2 Experimental

1M stoichiometric nitric solutions for LSM were prepared by dissolution of $Sr(NO_3)_2$, $La(NO_3)_3 \cdot 6H_2O$ and $C_4H_6MnO_4 \cdot 4H_2O$ in pure water by the Liquid Mix technique. The solution was de-hydrated with a 24 hours thermal treatment at 100°C. The resulting material was manually milled and then, de-nitrated at 500°C for 10 hours. The LSM compound was finally obtained by calcination at different synthesis temperatures (ST) to obtained several particle sizes in the precursors. We note this material as precursor LSM_LM.

Alternatively, another procedure was followed in order to obtain separated by adding of NaCl after the de-nitration process, preventing agglomeration in the precursor powder. It was added in a 1:4 relation and mixed with the de-nitrated powder in a ball miller for 1 hour. After that, a final thermal treatment was made at ST between 500ºC and 1000ºC. The NaCl was removed by several dissolution cycles in pure water. This material will be referred from now on as LSM_NC.

To reach the same objective, addition of Ammonium Nitrate was made in a 1:10 relation after milling. This materials were mixed in a ball miller for 1 hour. The Ammonium Nitrate evaporated during the calcination at ST = 700ºC. This material will be named LSM_AN.

X-ray powder diffraction (XRD) analysis was performed with a PANalytical Empyrean diffractometer with a PIXcel3Dt detector using Cu-Kα radiation ($\lambda = 1.54056$ Å). Scanning electronic microscopy (SEM) measurements were made using a SEM FEI Quanta 200 and a SEM Carl Zeiss NTS Supra 40.

Gadolinium-doped ceria (GDC) electrolytes were prepared from commercial $CeO_2$-20mol% $Gd_2O_3$ powders (NexTech Materials) by uniaxial pressing at 200 MPa and sintering at 1350°C for 2 h, obtaining 0.5 mm thick samples with a relative density higher than 96%.

All LSM precursor powders were made into an ink with commercial Decoflux$^{TM}$ (WB41 Zschimmer and Schwarz) polymerical solution in 1:2 (material:decoflux) mass proportion. The ink was smeared on the electrolyte as thinly as possible with a brush and dried at 50°C in air for about 20 min. Then, the cathodes were fired at temperatures (TT) between 1000°C and 1300°C for 1 h for the cathode-electrolyte sintering, in order to obtain cathodes formed by particles of different diameters.

The area specific resistance (ASR) of the cathodes was obtained in symmetrical [LSM/GDC/LSM] cells from electrochemical impedance spectroscopy (EIS) measurements, using a Gamry 750 potentiost-galvanostat-impedance analyzer with Ag paste as current collector. Measurements were performed at zero bias with an amplitude of 20 mV. The porcentual error for ASR reaches a maximum of 5%, much below the size of the points used to display the values in the figures.

The following paragraph summarizes the nomenclature of the materials used as precursors for the cathode and for the cathodes themselves.
Precursors:
- Synthesized by Liquid Mix: LSM_LM
- Synthesized by Liquid Mix with the adding of NaCl: LSM_NC
- Synthesized by Liquid Mix with the adding of Ammonium nitrate: LSM_AN
Cathodes will be named as "*synthesis procedure*"_"ST"_"TT". The *synthesis procedure* stands for LM, NC, AN for Liquid Mix, Liquid Mix+NaCl and Liquid Mix+Ammonium Nitrate, respectively. ST is the synthesis temperature and TT is the temperature of the thermal treatment used for cathode-electrolyte sintering.

## 3 Results and discussions

### 3.1 Liquid Mix Method

Figure 1 shows the X-ray diffractograms of LSM_LM powders used as precursors to make the cathodes, before being attached to the electrolyte. Samples synthesized at 600ºC and 1000ºC display a rhombohedral crystalline structure as previously reported [21], with no evidence of impurity phases. The crystallite sizes, obtained using the Scherrer equation, are 25±10 nm and 270±10 nm for the powders obtained at 600ºC and 1000ºC, respectively.

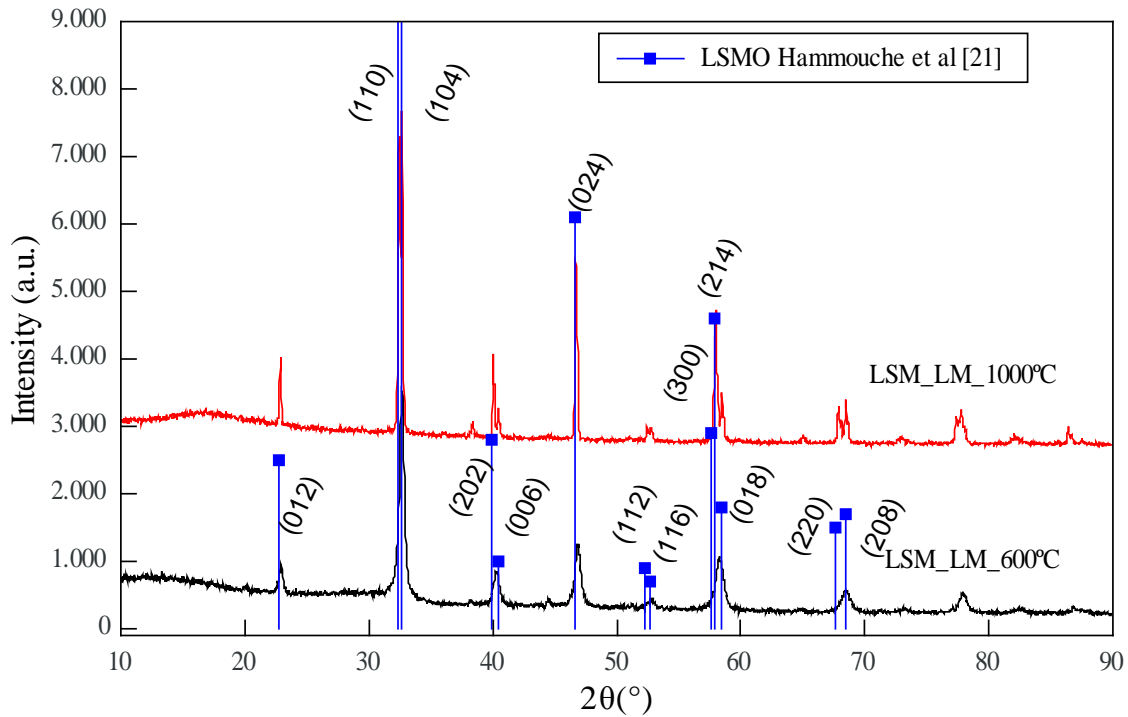

Figure 1: X-ray diffractograms of the LSM powders synthesized by Liquid Mix and treated at ST = 600ºC and 1000ºC. XRD reflections from reference [21] are indicated by points (color online).

Figure 2 presents SEM images of the precursor powders. It can be seen that the particles diameters are in the nanometric scale and that they are agglomerated in larger scale structures on the μm order. Those powders were used to make an ink and sintered on the electrolytes to obtain the cathodes in symmetrical cells.

Complementary, figure 3 shows SEM images of the surface of the different cathodes. An homogeneous distribution of the particles can be observed. A comparison of figures 3 (a) and (b), evidences that the LM_600_1000 and LM_1000_1000 cathodes (using powders synthesized at ST=600ºC and 1000ºC respectively and then both sintered at TT=1000ºC) are highly porous, however a much continuous layer is obtained using powders synthesized at ST=1000ºC (fig. 3(b)), as compared with those obtained at ST=600ºC (fig. 3(a)). By comparing the LM_1000_1000 and the LM_1000_1300 cathodes (figs 3(c) and (d)), it has seen somehow that porosity is lost when increasing the TT. Table I summarizes average particle diameter ($d$) of all cathodes, obtained from the analysis of the SEM images, which shows that the average particle diameter of particles that form the cathode is both determined by the synthesis temperature of the compound and the sintering temperature of the cathode, expressing $d$ ranges from 90 to around 2000 nm.

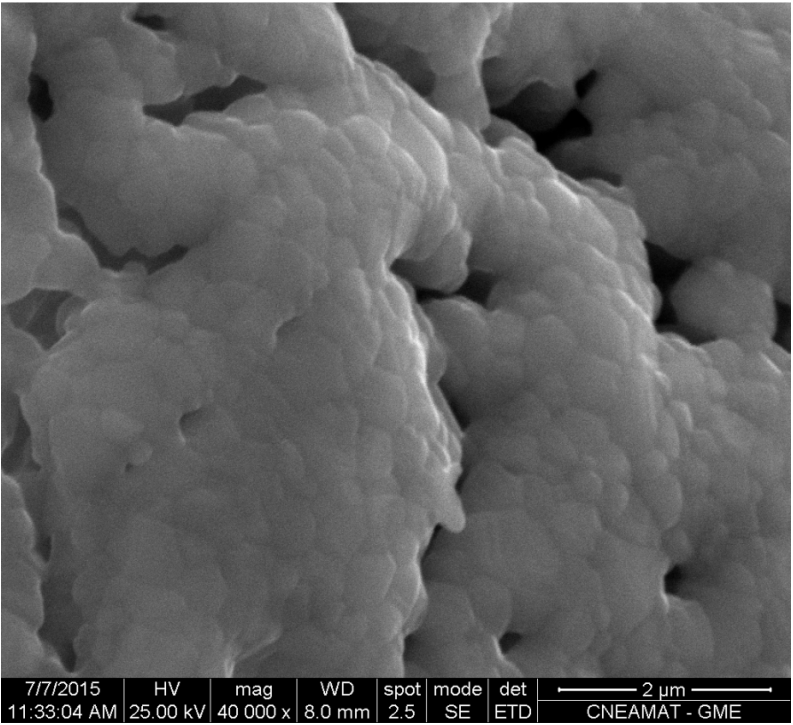
Figure 2: SEM images of the precursor powders obtained by Liquid Mix.

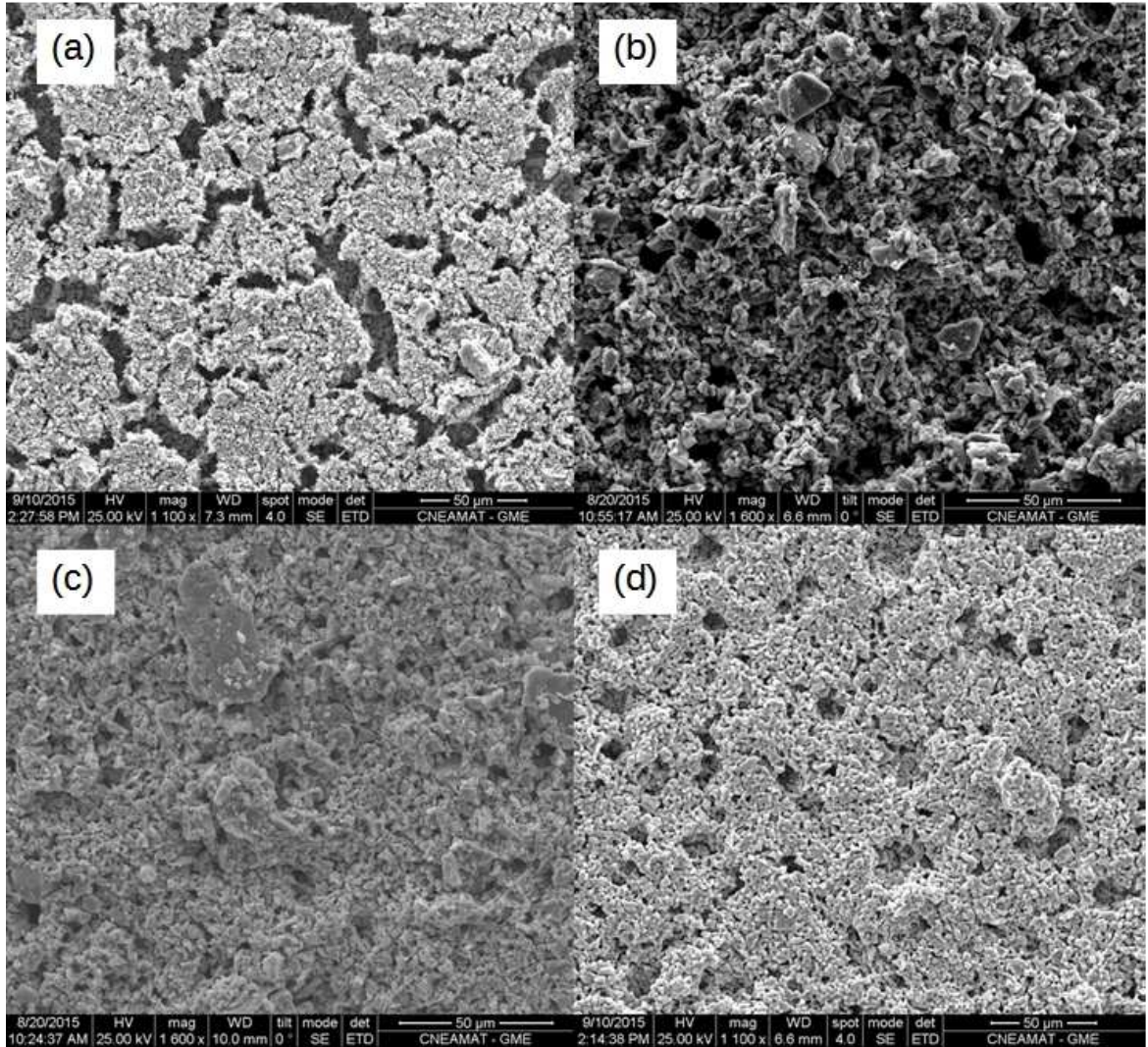

Figure 3: SEM micrographs of various cathodes obtained using Liquid Mix precursors. (a) LM_600_1000, (b) LM_1000_1000, (c) LM_1000_1100 and (d) LM_1000_1300.

|            | LM_600_1000 | LM_1000_1000 | LM_1000_1100 | LM_1000_1300 |
|------------|-------------|--------------|--------------|--------------|
| $d$ (nm)   | 90±5        | 300±15       | 375±20       | 2340±80      |

**Table I:** Average particle diameter ($d$) for the cathodes made with powders synthesized by Liquid Mix. Cathodes developed with powders synthesized at ST of 600ºC and 1000ºC and sintered at TT of 1000ºC, 1100ºC and 1300ºC are presented.

Figure 4 represents the temperature dependence of Area Specific Resistance (ASR), obtained from the impedance spectra of the cathodes. The lowest ASR value is obtained for the LM_1000_1300 cathode (i.e. the one formed by particles of largest diameter), while the other three display similar values in almost the whole temperature range. However, the ASR values corresponding to the different cathodes, seem to practically merge at 700ºC.
The dependence of the ASR with the average diameter of the particles, for cathodes obtained using Liquid Mix powders is presented in figure 5 for selected temperatures. At around 500ºC, cathodes made with larger diameter particles clearly display the lowest ASR, while at larger measuring temperatures (~700ºC) this difference, although present, is less clear.

While analyzing the previous results, it has to be considered that even though we have obtained cathodes formed with particles of sizes ranging between ~90 nm and ~ 2000 nm, the surface of those particles is not largely exposed to the atmosphere, as evidenced by the high degree of agglomeration shown in figure 2. Thus, the main effect dominating the performance of the present cathodes is expected to be the quality of the attachment between cathode and electrolyte. In fact, the cathode with the best performance is the one that has been sintered at larger temperature, supporting that picture.

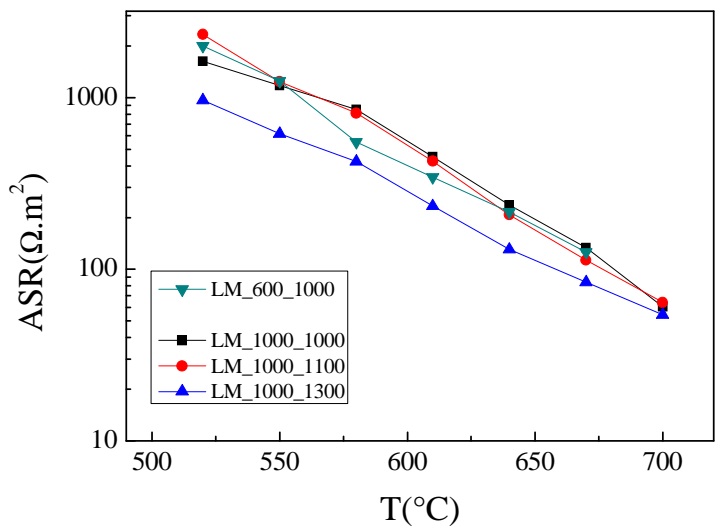

Figure 4: Area Specific Resistance as a function of temperature for LM_600_1000, LM_1000_1000, LM_1000_1100 and LM_1000_1300 cathodes. Cathodes are presented according the synthesis temperature (ST) and thermal treatment sintering temperature (TT) as LM_"ST"_"TT".

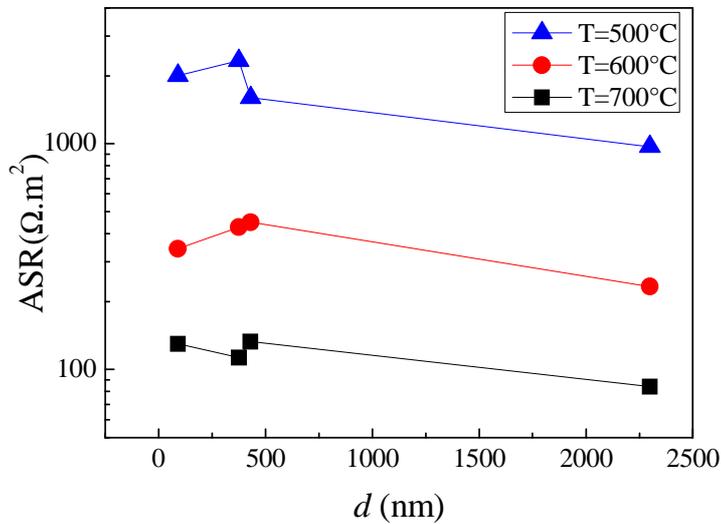

Figure 5: Area Specific Resistance as a function of the average diameter of the particles that form the cathode (*d*) for cathodes obtained by the Liquid Mix method.

In any case, as all cathodes arise from agglomerated particles, proper analysis of the size effect on the ASR for this synthesis procedure cannot be made. For that reason, the following sections presents two procedures to obtain cathodes in which the particles remain deagglomerated.

**3.2 Using NaCl as a deagglomerating Agent**

Figure 6 shows the X-ray diffraction patterns for the powders used as precursors of the cathode in which NaCl was used to inhibit the agglomeration of the particles. Powders synthesized at 500ºC, 700ºC and 1000ºC have been obtained. No evidence of segregated NaCl is observed. However, for 2θ > 50º, a slight shift towards larger angles is observed, suggesting a relative shrink of the unit cell.

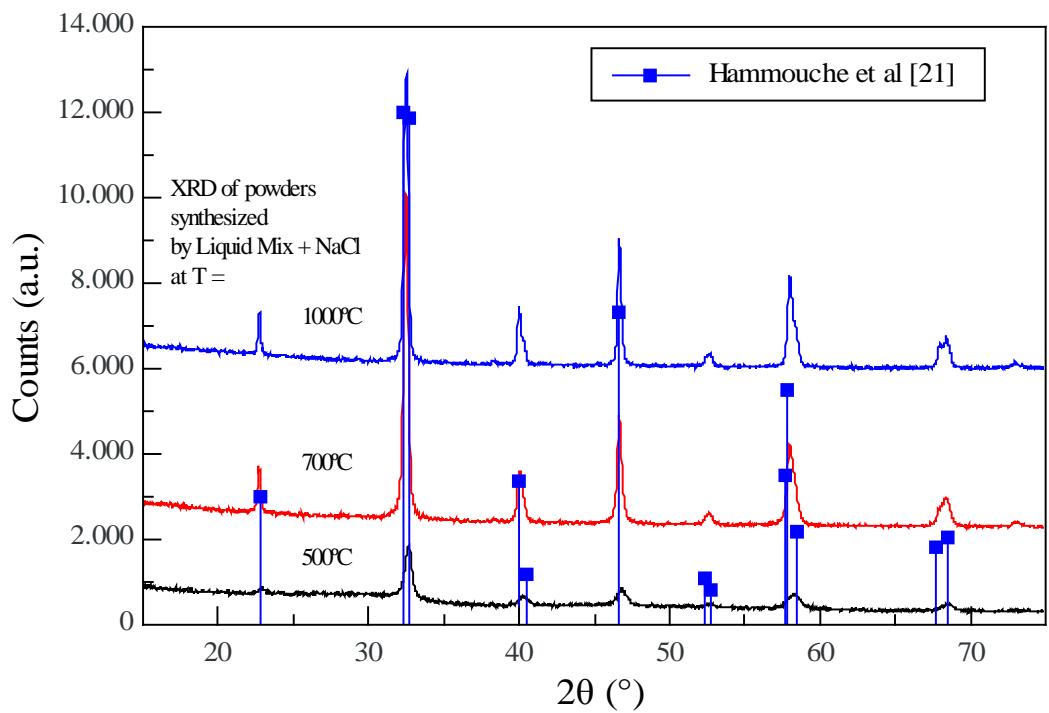

Figure 6: X-ray diffractograms of the LSM powders synthesized by Liquid Mix with NaCl as deagglomerating agent, calcined at ST = 500ºC, 700ºC and 1000ºC. XRD reflections from reference [21] are indicated by points (color online).

Extreme care was taken in order to dissolve the NaCl. However, EDS measurements show a minor proportion of around 1% of Cl and Na in the powder synthesized at 700ºC and 1% of Na in the powder synthesized at 1000ºC.

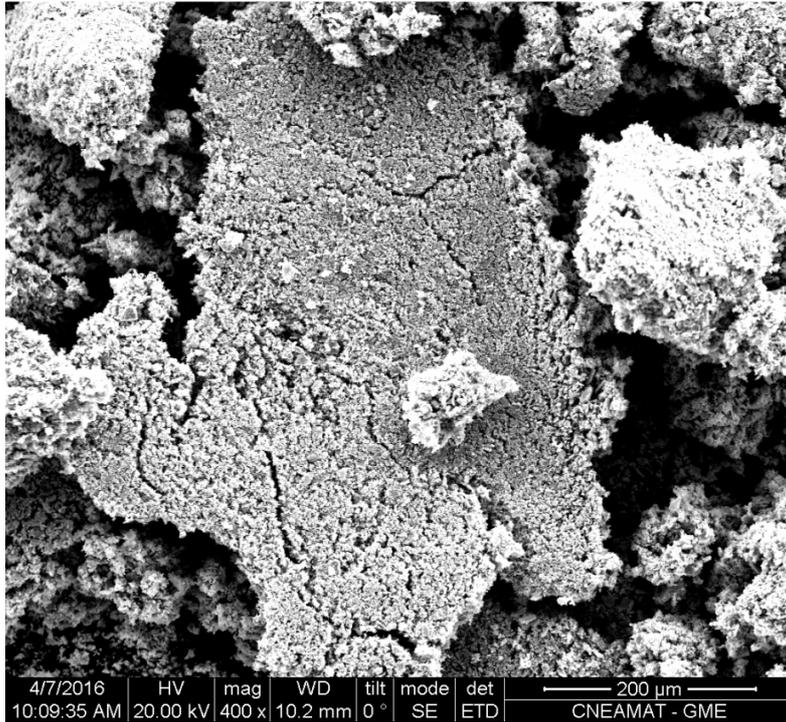

Figure 7: SEM images of the precursor powders obtained by Liquid Mix with NaCl (LM_NC) as deagglomerating agent.

Figure 7 shows the SEM micrograph of the LSM_NC powder used as precursor for the cathodes, synthesized at 500ºC, where it can be seen that particles are dispersed.

SEM micrographs of the cathodes are shown in figure 8. It is clear that the particles that form the cathodes are separated among them, in contrast with cathodes obtained by liquid mix (see figure 3 of Section 3.1). In figure 8 (a), (b) and (c), morphology of cathodes made with precursors synthesized at different temperatures (ST=500ºC, 700ºC and 1000ºC) but sintered at the same temperature (TT=1000ºC) can be compared. Particle diameters that form the cathodes are presented in Table II. The diameter of the particles presents a slight dependence with ST, as can be seen by comparing the values of *d* of the sintered cathodes at TT=1000ºC. There is a small increase in the diameter of the particles from the NC_500_1000 to the NC_1000_1000 cathode, a fact that can be expected as the last cathode was subjected twice to a thermal treatment at 1000°C.

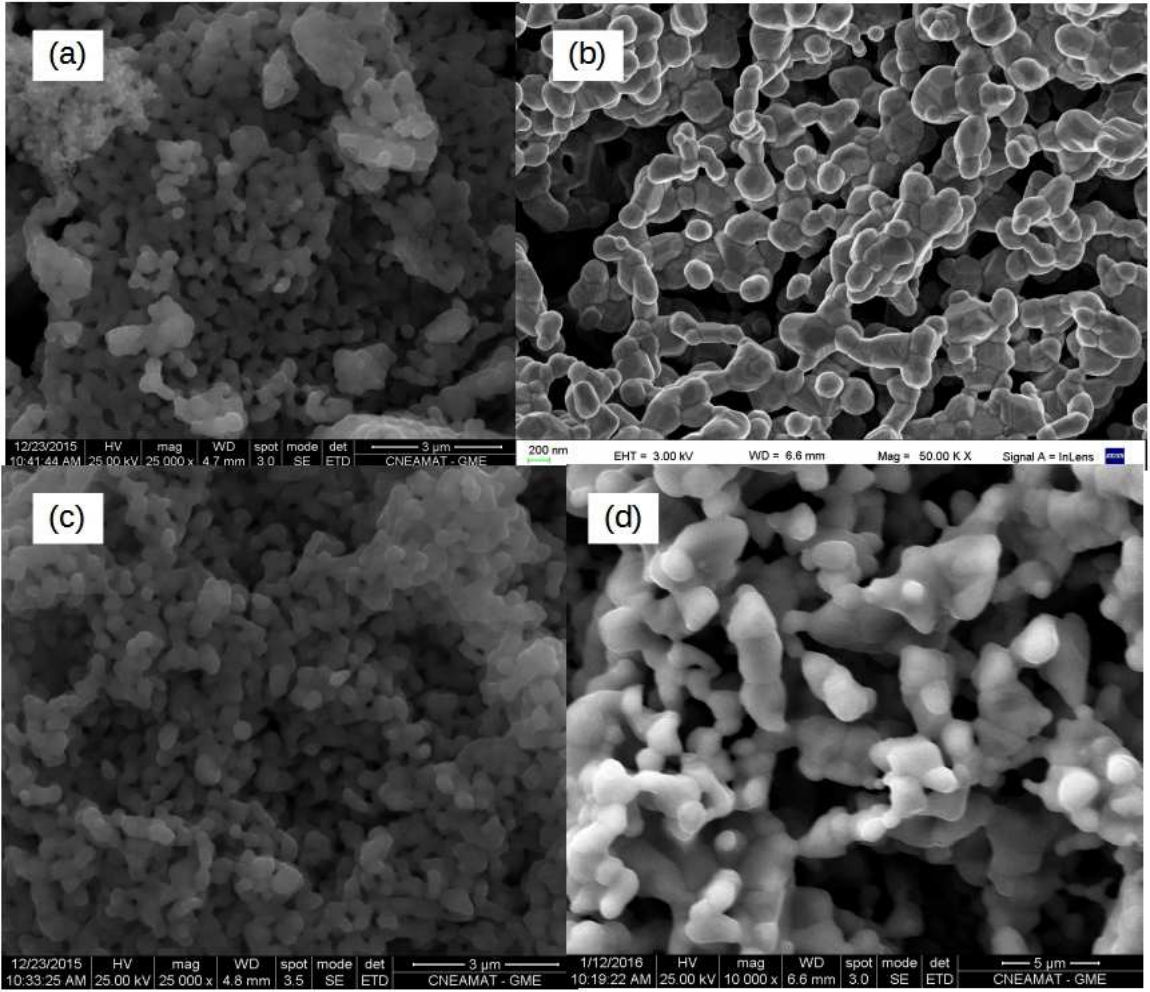

Figure 8: SEM micrographs of various cathodes obtained using precursors with the adding of NaCl. (a) NC_500_1000, (b) NC_700_1000, (c) NC_1000_1000 and (d) NC_1000_1300.

Figure 8 (d) shows the SEM micrograph of the surface of the NC_1000_1300 cathode. A comparison with the NC_1000_1000 cathode proves that increasing the temperature of the cathode sintering procedure, provokes a clear enlargement of one order of magnitude in the diameter of the particles that form the cathode. Table II presents the average particle diameter corresponding to all cathodes, showing that they range between 200 to 2000 nm.

|  | NC_500_1000 | NC_700_1000 | NC_700_1100 | NC_1000_1000 | NC_1000_1300 |
|---|---|---|---|---|---|
| $d$ (nm) | 230±20 | 230±20 | 450±20 | 270±20 | 2000±200 |

**Table II:** Average particle diameter for the cathodes made with powders synthesized using NaCl as a deagglomerating agent.

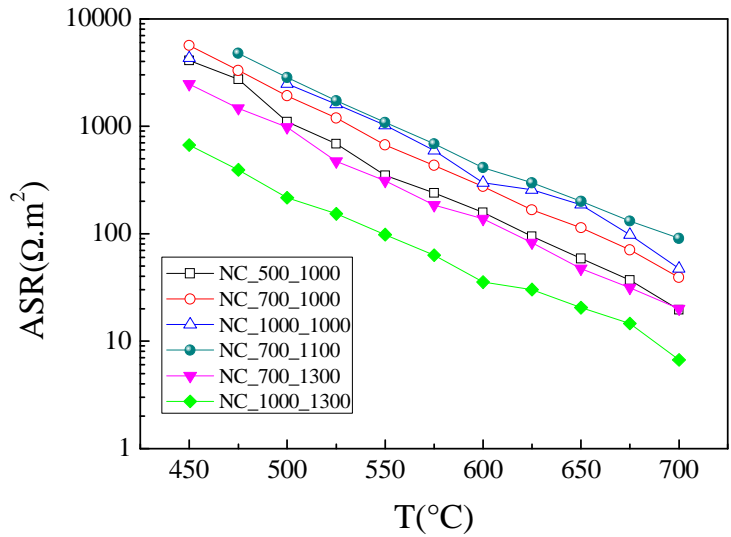

Figure 9: Area Specific Resistance as a function of temperature for cathodes obtained with precursors deagglomerated using NaCl. Cathodes are presented according the synthesis temperature (ST) and thermal treatment sintering temperature (TT) as NC_"ST"_"TT" (color online).

Figure 9 presents the ASR as a function of temperature for selected cathodes. The most significant feature is that the cathode with the lowest ASR in the whole temperature range, is the one synthesized and sintered at the largest temperatures. We further analyze the ASR values measured at 650ºC, because it is representative of the intermediate temperature range of operation of SOFC, which is displayed in figure 10. Figure 10(a) shows the ASR(650ºC) of the NC_ST_1000 cathodes as a function of the synthesis temperature. Even though particle diameters in these cathodes do not display a very large change, it is expected that the powders used as precursors should be formed by particles which size is determined by the ST. In fact, a small increase can be observed in Table II when rising ST from 500ºC to 1000ºC, as evidenced by the comparison of the values of *d* for the NC_500_1000 and NC_1000_1000. At the same time, there is a clear increase in the ASR values when rising ST, which indicates that the use of smaller nanoparticles as a starting point of the cathode is beneficial for its performance.

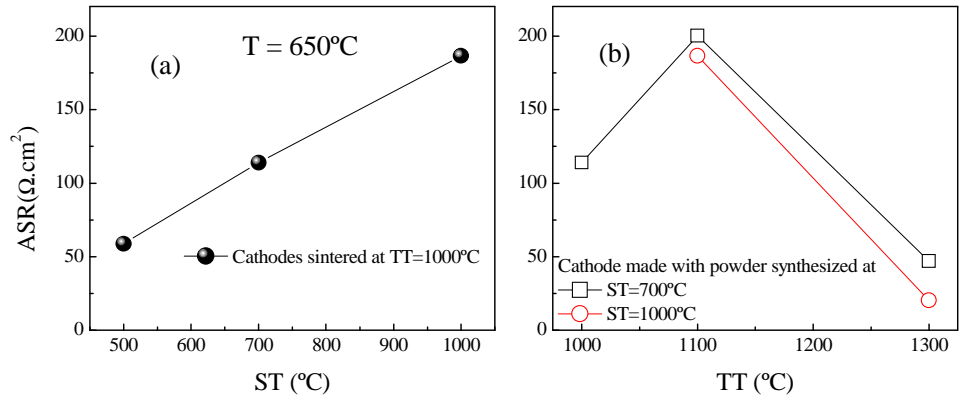

Figure 10: Area Specific Resistance measured at 650ºC: as a function of (a) the synthesis temperature (ST) and (b) the thermal treatment sintering temperature (TT), for cathodes obtained with precursors deagglomerated using NaCl (color online).

The dependence of the ASR with TT are shown in figure 10 (b), for cathodes made by using precursors synthered at 700°C and 1000°C. In both cases, the cathodes with TT = 1100ºC display larger ASR than the cathode with TT = 1300ºC. However, we can see that ASR can be reduced on increasing TT from 1100ºC to 1300ºC and also on reducing TT from 1100ºC to 1000ºC (only shown for the cathode with ST=700ºC). This non-monotonous behavior suggests that the ASR is governed by two regimes limited here by TT=1100ºC. To gain further insight into this phenomenon, it has been studied the value of the ASR at selected temperatures between 500ºC and 700ºC for all cathodes of this method, using the diameter of the particles "$d$" as parameter. In figure 11 it can be clearly seen that the alluded two-regime behavior is not a particularity of one cathode, but a general trend of all cathodes obtained through the present procedure. Thus, the two regimes can be better described by characterized by the "larger" and "smaller" particles that form the cathodes, where the separation between them is around a particle diameter of 500 nm.

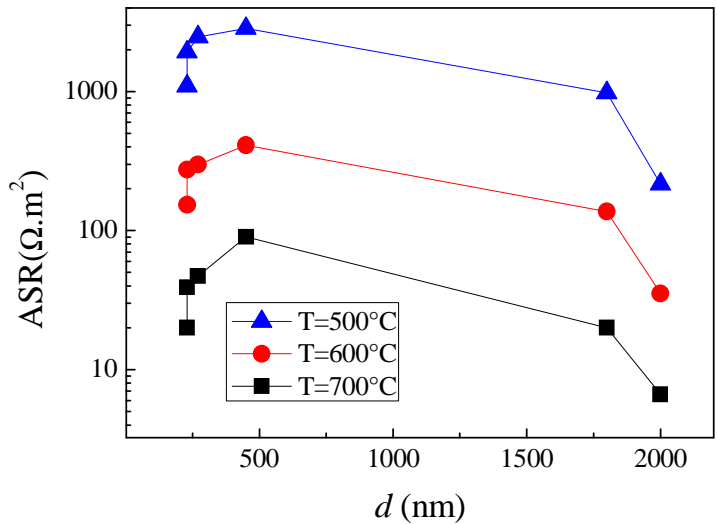

Figure 11: Area Specific Resistance as a function of the average diameter of the particles that form the cathode (*d*) for cathodes obtained by the Liquid Mix method using NaCl as deagglomerating agent (color online).

**3.3 Using Ammonium Nitrate as a deagglomerating Agent**

The previous method (Using NaCl as a deagglomerating Agent), was used to obtain deagglomerated particles. Also, it has been explored an alternative method to obtain disperse nanoparticles but using an organic compound as a deagglomerating agent. Ammonium Nitrate was chosen because this compound can be completely removed after calcination at 700ºC. Figure 12 shows the X-ray diffraction patterns for the powders used as precursors of the cathode in which Ammonium Nitrate was used to inhibit the agglomeration of the particles. As a result, reflections corresponding to $Mn_2O_3$ were further reduced in its concentration at around 1% after performing the sintering of the cathodes.

Figure 13 represents a SEM micrograph of the precursor material, from which is clear that the dispersion of the particles has been achieved.

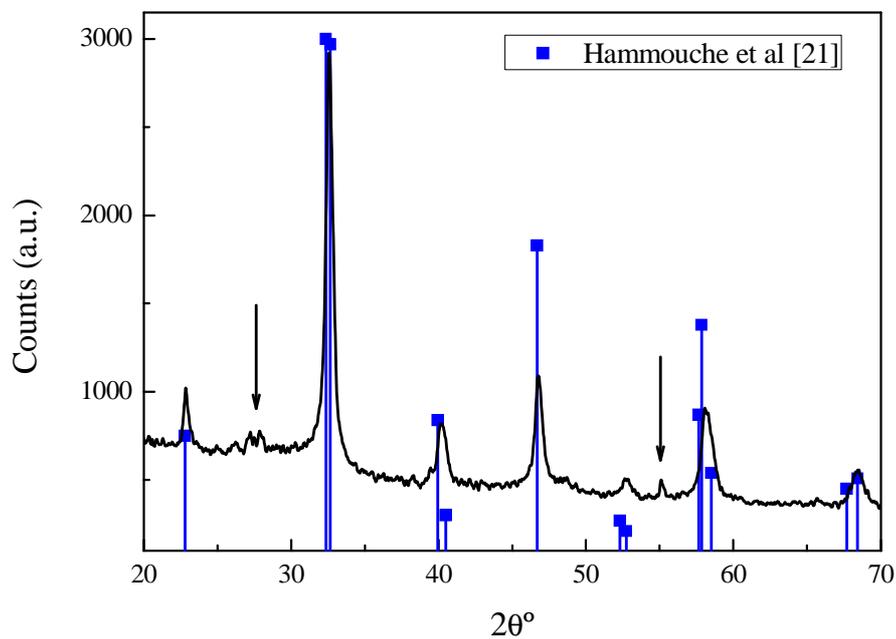

Figure 12: X-ray diffractograms of the LSM powders synthesized by Liquid Mix with Ammonium Nitrate as deagglomerating agent. XRD reflections from reference [21] are indicated by points (color online).

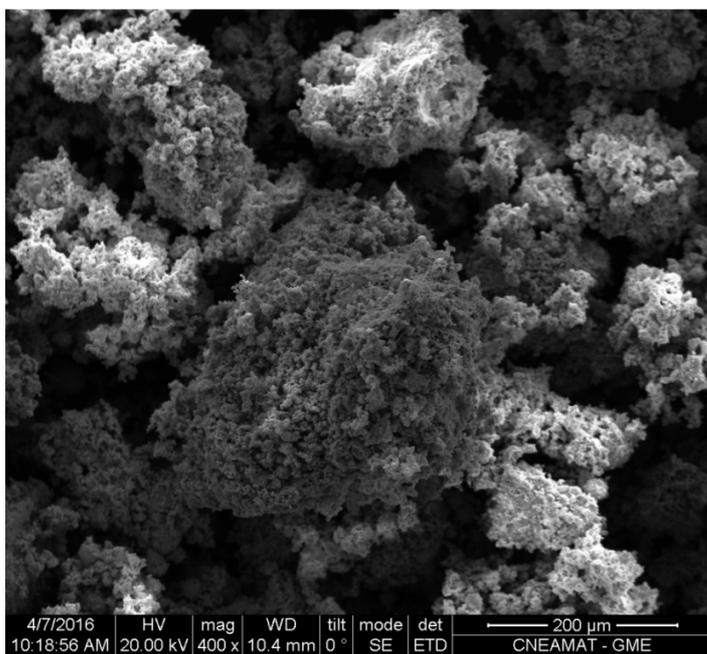

Figure 13: SEM images of the precursor powders obtained by Liquid Mix with Ammonium Nitrate as deagglomerating agent.

Cathodes were sintered at temperatures between 1000 and 1300ºC in order to obtain electrodes with different particle diameter. Regarding the morphology, the agglomeration of grains in the cathodes depends on the temperature of the sintering thermal treatment. In figure 14, it can be clearly seen that AN_700_1000 cathode is the one in which particles are comparatively much dispersed, and agglomeration increases towards the AN_700_1300 cathode. Table III presents the average grain sizes of the particles that form the LSM_AN cathodes, which range from 90 to 1200 nm.

|  | AN_700_1000 | AN_700_1100 | AN_700_1200 | AN_700_1300 |
|---|---|---|---|---|
| *d* (nm) | 90±20 | 310±20 | 680±20 | 1200±200 |

**Table III:** Average particle diameter for the cathodes made with powders synthesized using Ammonium Nitrate as a deagglomerating agent.

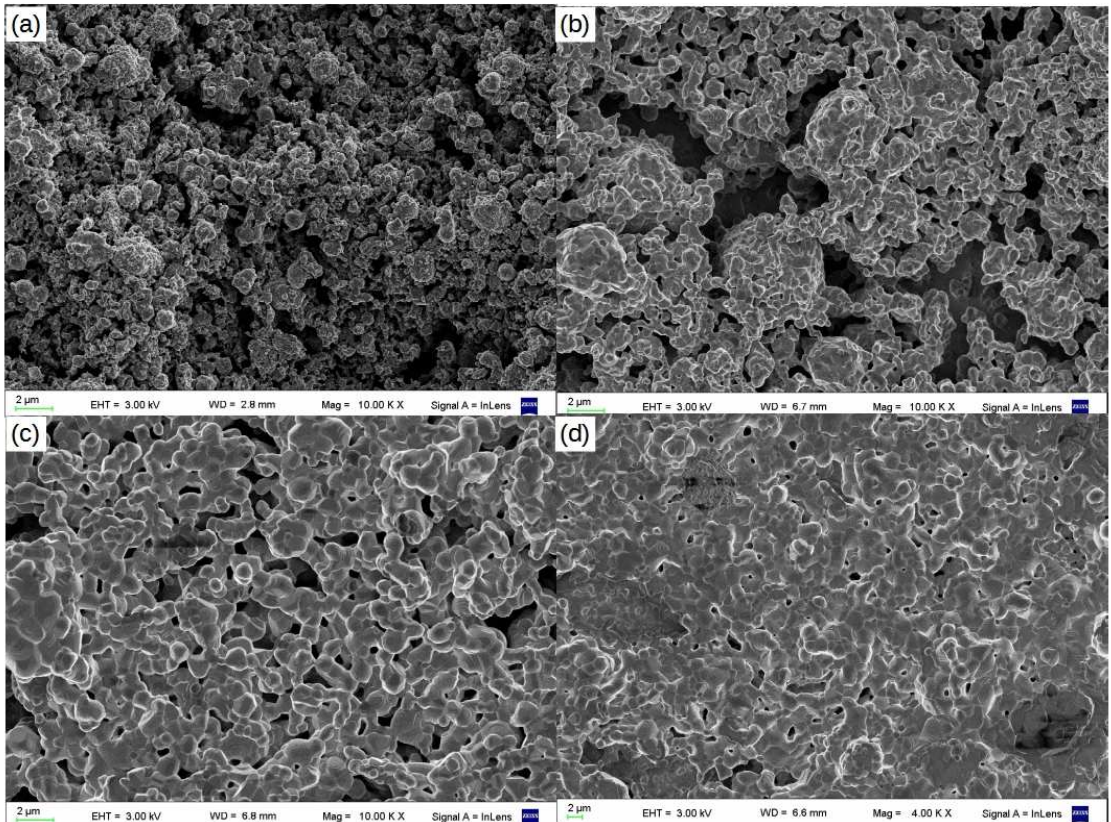

Figure 14: SEM micrographs of various cathodes obtained using precursors with the add of Ammonium Nitrate. (a) AN_700_1000, (b) AN_700_1100, (c) AN_700_1200 and (d) AN_700_1300.

In figure 15, the ASR as a function of temperature is presented for all cathodes. The figure also presents the ASR of the cathodes as a function of the particle size for selected temperatures in figure 16, in which is shown that no monotonous dependence is observed. However, the cathode with larger ASR is the one formed with particles of smaller diameter. As a significant feature, the ASR values obtained by this method are the smallest among the three presented chemical routes.

The ammonium nitrate helped to obtain a disperse powder and in turn a cathode in which agglomeration can be controlled by the TT and showing the best performance. As a counterpart, the presence of a minor intermediate compound, identified as $Mn_2O_3$ has been detected.

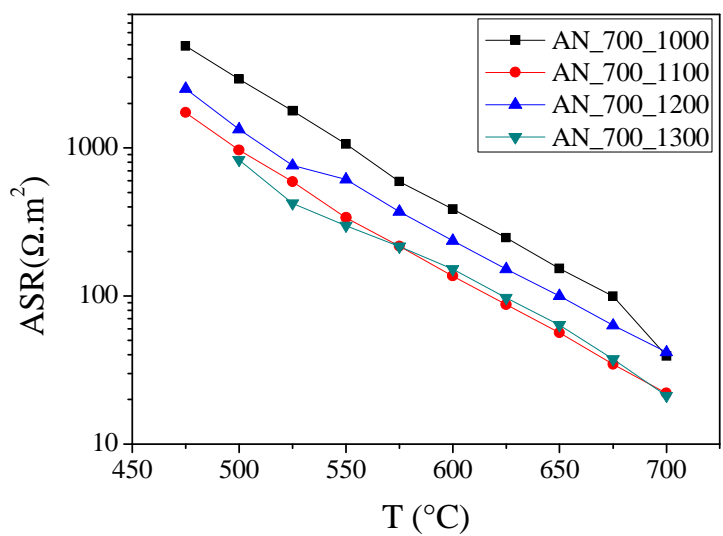

Figure 15: Area Specific Resistance as a function of temperature for cathodes obtained with precursors deagglomerated using Ammonium Nitrate. Cathodes are presented according the synthesis temperature (ST, here fixed at 700ºC) and thermal treatment sintering temperature (TT) as AN_(ST)_(TT) (color online).

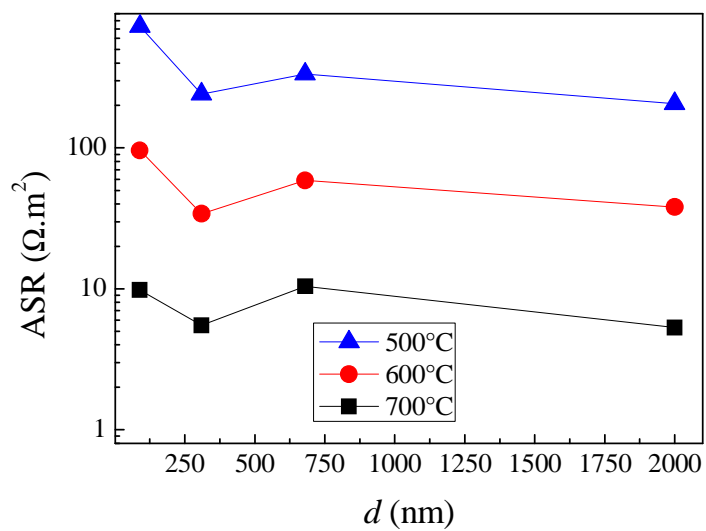

Figure 16: Area Specific Resistance as a function of the average diameter of the particles that form the cathode ($d$) for cathodes obtained by the Liquid Mix method using Ammonium Nitrate as deagglomerating agent (color online).

### 3.4 Comparison between methods

This section summarizes and compares the results obtained from all methods to gain insight that can be used in future studies or for applications.

Results show that cathodes formed by larger particles presented the lowest ASR for all methods. Figure 17 presents the ASR at 700ºC separated in three ranges of particle diameter, as a function of the synthesis method. As stated before, cathodes formed by larger particles display the better performance for the three methods showing that both methods used to deagglomerate the particles, results in a comparative smaller ASR. The (Liquid Mix + Ammonium Nitrate) is the best method in this aspect.

The results presented for cathodes using NaCl (fig. 10(a)), suggests that the use of smaller particles in the precursor materials as a starting point is beneficial for the cathodes performance.

The performance dependence with the particle size of the cathodes is different in each method, thus not allowing to draw a single conclusion for all methods. However, we can see that in the (Liquid Mix + Ammonium Nitrate) method, which is the one displaying an optimized cathodic performance, the dependence with $d$ is less pronounced as compared with the other two methods.

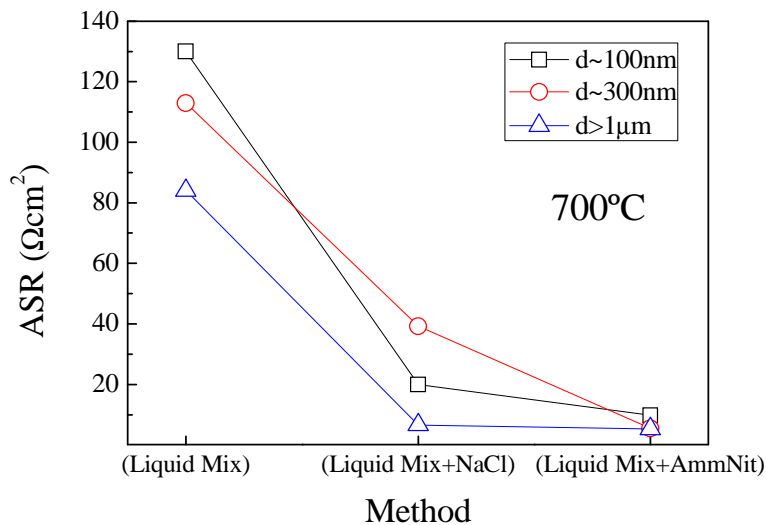

Figure 17: Area Specific Resistance at 700ºC for the different methods of synthesis separated by three ranges of $d$: "small"~100nm, "medium"~300nm and "large"~1µm (color online).

### 4 Conclusions

In summary, the dependence of the SOFC cathodic performance of LSM nanoparticles has been studied, focusing on two factors: the degree of agglomeration of the particles and the particle diameter.

The results allows to conclude that deagglomeration of the particles is beneficial for cathodic performance, as the largest ASR was shown by the cathode synthesized by Liquid Mix, which is formed by highly agglomerated particles. Thus, precursors in which particles are separated should be used in order to prevent agglomeration during the cathode-electrolyte sintering. The enhancement achieved by deagglomeration can even overcome the eventual negative effect of a small amount of undesired impurities, as also shown by the experiments.

As for the synthesis method, the cathode displaying the best performance is the one made with particles obtained by the Liquid Mix method, adding Ammonium Nitrate as a deagglomerating agent. However, as minor impurities identified as $Mn_2O_3$ were observed, further work is needed in order to improve this method.

Regarding the effect of particle diameter, no clear dependence has been observed. The cathodes with larger particle sizes are those with the best performance, apparently in contradiction with previous works in which nanostructuring is shown to improve cathodic performance for other materials such as cobaltites. However, the larger particles are obtained by comparatively higher temperatures for the thermal treatment for cathode adherence. Thus, and also considering that this trend was observed for all three methods, even in the one in which particles are highly agglomerated, we believe that the responsible for the optimized behavior is not the size of the particles, but the better cathode-electrolyte adherence. In any case, it is clear that much work is needed to elucidate this point.

The overall conclusion is that the best method to obtain LSM nanostructured cathodes is the (Liquid Mix + Ammonium Nitrate), with a cathode sintering at TT ~ 1300ºC. However, the dependence with the sintering temperature is less significant in this method than in the other two.

## 5 References

[1] B.C.H. Steele, A. Heinzel, Nature 414 (2001) 345.
[2] S. Kartha, P. Grimes, Phys. Today 47 (1994) 54–61.
[3] S.C. Singhal, K. Kendal, High-Temperature Solid Oxide Fuel Cells: Fundamentals, Design and Applications, Elsevier, 2003
[4] J.H. Choi, J.H. Jang, J.H. Ryu, S.M. Oh. J. Power Sources 87 (2000) 92–100.
[5] S.P. Yoon, J. Han, S.W. Nam, T.-H. Lim, I.-H. Oh, S.-A. Hong, Y.-S. Yoo, H.C. Lim. Performance of anode-supported solid oxide fuel cell with $La_{0.85}Sr_{0.15}MnO_3$ cathode modified by sol−gel coating technique. J. Power Sources 106 (2002) 160–166.
[6] T. Suzuki, M. Awano, P. Jasinski, V. Petrovsky, H.U. Anderson. Solid State Ionics 177 (2006) 2071–2074.
[7] V. A. C. Haanappel, J. Mertens, D. Rutenbeck, C. Tropartz, W. Herzhof, D. Sebold, F. Tietz. J. Power Sources 141 (2005) 216–226.
[8] Y. Takeda, R. Kanno, M. Noda, O. Yamamoto, Bull. Inst. Chem. Res. 64 (1986) 157.
[9] H. Uchida, S. Arisaka, M. Watanabe, Solid State Ionics 135 (2000) 347–351.
[10] J. Sacanell et al. Physica B 398 (2007) 341–343
[11] L. Baqué, A. Caneiro, M.S. Moreno, A. Serquis. Electrochem. Commun. 10 (2008) 1905–1908.
[12] J. Yoon, S. Cho, J.-H. Kim, J.H. Lee, Z. Bi, A. Serquis, X. Zhang, A. Manthiram, H. Wang. Adv. Funct. Mater. 19 (2009) 3868−3873.

[13] J. Sacanell, A.G. Leyva, M. Bellino, D.G. Lamas. J. Power Sources 195 (2010) 1786–1792.
[14] A. Mejía Gómez, J. Sacanell, A.G. Leyva, D.G. Lamas. Ceram. Int. 42 (2016) 3145–3153.
[15] K. Bentley, J.S. Trethewey, A.B. Ellis, W.C. Crone, Nano Lett. 4 (2004) 487
[16] L. Hueso, N. Mathur. Nature 427 (2004) 301–304.
[17] A.V. Berenov, J.L. MacManus-Driscoll, J.A. Kilner. Solid State Ionics 122 (1999) 41–49.
[18] R.A. De Souza, J.A. Kilner, J.F. Walker. Mater. Lett. 43 (2000) 43–52.
[19] E. Navickas, T. Huber, Y. Chen, W. Hetaba, G. Holzlechner, G. Rupp, M. Stöger-Pollach, G. Friedbacher, H. Hutter, B. Yildiz, J. Fleig. Phys. Chem. Chem. Phys. 17 (2015) 7659–7669.
[20] I. Rossetti, M. Allieta, C. Biffi, M. Scavini. Phys. Chem. Chem. Phys. 15 (2013) 16779–16787.
[21] A. Hammouche, E. Siebert, and M. Kleitz, Electrochimica Acta 40 (1995) 1741
[22] Joaquín Sacanell, Joaquín Hernández Sánchez, Adrián Ezequiel Rubio López, Hernán Martinelli, Jimena Siepe, Ana G. Leyva, Valeria Ferrari, Dilson Juan, Miguel Pruneda, Augusto Mejía Gómez, and Diego G. Lamas. J. Phys. Chem. C 121 (2017) 6533−6539. DOI: 10.1021/acs.jpcc.7b00627